\documentclass{aa}                     

\usepackage{graphicx} 

\begin{document}

 \title{Implications of Cosmological Gamma-Ray Absorption} 
\subtitle{I. Evolution of the Metagalactic Radiation Field}

 \author{T. M. Kneiske\inst{1} \and K. Mannheim\inst{1}
 \and D. H. Hartmann\inst{2}} 

 \institute{Universit\"at W\"urzburg, Am Hubland, 97057 W\"urzburg, Germany
 \and
 Clemson University, Clemson, SC 29634-0978, USA}
 \offprints{T. M. Kneiske
 \email{kneiske@astro.uni-wuerzburg.de}}
 
 \date{Received/Accepted}

 \abstract{                                 
 Gamma-ray absorption due to  $\gamma\gamma$-pair creation on cosmological scales 
 depends on the line-of-sight integral of the evolving density of low-energy photons
 in the Universe, i.e. on the history of the diffuse, isotropic
 radiation field.
 Here we present and discuss a semi-empirical model for this metagalactic 
 radiation field based on stellar light produced and reprocessed in evolving galaxies. 
 With a minimum of parameters and assumptions, 
 the present-day background 
 intensity is obtained from the far-IR to the ultraviolet band.
 Predicted model intensities are independent of cosmological parameters, since we require 
 that the comoving emissivity, as a function of redshift, 
 agrees with observed values obtained from deep galaxy surveys.
The far-infrared background at present day prediced from optical galaxy surveys
falls short in explaining the observed one, and we show that this deficit can be removed
by taking into account dusty galaxies with a seperate star formation rate.
The accuracy and reliability of the model, 
out to redshifts of $z\sim 5$, allow a realistic estimate of the 
attenuation length of GeV-to-TeV gamma-rays and its uncertainty, which is
the focus of a subsequent paper. 
}

\maketitle
 
 \section{Introduction}
 
 Understanding the evolution of large scale structure in the Universe is a major 
 goal of modern observational cosmology. Numerical simulations of hierarchical
 structure formation in a globally homogeneous universe are now tractable
 (e.g., Nagamine, Cen, $\&$ Ostriker 2000; Kauffmann et al. 1999), but
 connecting the evolving structures to observable
 fluxes of electromagnetic radiation involves uncertain empirical descriptions of
 star formation, supernova feedback, and the dust-gas interplay. The necessary input 
 comes from extensive observational campaigns, such as deep galaxy surveys, which 
 measure the number of galaxies, their morphological types, colors, fluxes, and distances
 in presumably representative solid angles out to redshifts of z $\sim 6$. The wealth
 of detailed information derived from these observations can significantly complicate
 the effort to link theories of galaxy evolution and large scale structure formation.
 It is helpful to single out global quantities for which predictions can be compared with
 observations. One such quantity is the cosmic star formation rate (SFR) and its 
 associated metagalactic radiation field (MRF) 
 (the MRF at $z=0$ is commonly referred to as Extragalactic Background Light, EBL). 
 The contribution of galaxies to the 
 MRF is most significant between the far-infrared and the ultraviolet, while at longer 
 wavelengths cosmic 2.7~K microwave background (CMB) radiation from the big bang dominates. 
 At shorter wavelengths, accretion-powered active galactic nuclei provide much of the 
 high-energy background (e.g., Mushotzky et al. 2000; Sreekumar 2000). In the GeV
 regime the MRF seems to originate from unresolved blazars and galaxies.
 Gamma rays emitted by novae, supernovae, and $\gamma$-ray bursts contribute the bulk of the 
 observed background in the window around 1 MeV (e.g., Watanabe et al. 1999; 
 Weidenspointner 1999, Ruiz-Lapuente et al. 2000). 
 In principle, the evolution of the MRF should be predictable with structure formation
 models (e.g., Sommerville \& Primack 1999), so that the observed MRF 
 could be used to infer either the role of AGNs, low surface brightness objects, decays
 of relic particles, or to single out cosmological parameters.
 However, these models still rely on a wealth of uncertain parameters, and
 we are far from the ultimate goal of a first principles theory of the MRF.
For an overview sie Hauser \& Dwek (2001) and references therein.
 
 Three basic methods are commonly employed for computations of the MRF from luminosity
 functions undergoing (i) forward evolution from a theoretically determined initial
 state, (ii) backward evolution from an observationally given final (present-day) 
 state, or (iii) evolution that is directly observed over some range in redshift. 
 
 Method (i) starts from the theoretical framework of structure formation and
 evolution and predicts how luminosity functions evolve forward in cosmic
 time. The semi-analytic models of galaxy formation (e.g., White $\&$ Frenk 1991;
 Baugh et al. 1998; Kauffmann, White, $\&$ Guiderdoni 1993; Sommerville $\&$ 
 Primack 1999, Granato et al. 2000) are based on structure formation studies
 with dissipationless N-body simulations. These studies yield luminosity functions 
 for different morphological galaxy types that are
 in reasonable agreement with the observations. However, regarding the MRF,
 predictions from these models generally fail to satisfactorily reproduce observed
 cosmic emissivities. To better match these emissivities the models often 
 require significant adjustments in the prescriptions of star formation, supernova
 feedback, and the inclusion of further astrophysical
 effects which presently can not be calculated from first principles.
 A simplified model for the infrared and sub-mm range was developed by
 Guiderdoni et al. (1998), while Malkan \& Stecker (1998, 2001), en route of method (ii),
 determine the infrared MRF from local luminosity functions obtained with IRAS. 
 In a very substantial paper, Franceschini et al. (2001) employ recent ISO data 
 and more detailed models for the IR emission.\\

 Method (iii) computes the MRF directly from the global SFR inferred from
 tracers of cosmic chemical evolution, such as the various Lyman $\alpha$ systems 
 (Salamon \& Stecker 1998, Pei, Fall \& Hauser 1998), or from deep galaxy surveys
 (e.g., Madau et al. 1998, Rowan-Robinson 2001, Franceschini 2000).
 The spectral energy distribution (SED) for the globally averaged stellar population
 residing in galaxies can be estimated with population synthesis models (e.g. Bruzual 
 \& Charlot 1993) available for various input parameters, of which the
 initial mass function (IMF) and metallicity are the most important ones. 
 Reprocessing by gas and dust can be taken into account explicitly via some 
 model of the evolution of the dust and gas content in galaxies, in combination
 with assumed dust properties derived from local observations in the Milky Way.\\
  
 Observational attempts to determine or constrain the present-day background face  
 severe problems due to emissions from the Galaxy, which can introduce 
 large systematic errors. Nevertheless, a number of studies with COBE FIRAS 
 (Fixsen et al. 1998)
 and COBE DIRBE (Hauser et al. 1998) have resulted in highly significant 
 detections of a residual diffuse IR background,
 providing an upper bound on the MRF in the IR regime. Similarly, 
 the cumulative flux from galaxies detected in deep HST or 
 ISO exposures provide lower limits to the present-day MRF.  
 In the UV, measurements of the proximity effect provide an estimate of the MRF 
 at high redshifts (e.g., Giallongo et al. 1996).
 To constrain cosmic evolution of the MRF one can also utilize the fact that high-energy 
 gamma rays (from blazars or gamma-ray bursts) originating at large redshifts are
 attenuated by pair creation from interactions with low-energy MRF photons
 (e.g. Stanev\& Franceschini 1998, Renault et al. 2001), which is the subject of a subsequent publication
 in this series.\\

 Here we discuss a model of the evolving MRF that is based directly on observed
 emissivities (method iii), and is designed to use a minimal set of assumptions to 
 clearly reveal the connections between 
 input physics and output MRF. The method employed here (described in $\S$2) is  
 similar to the method discussed by Madau et al. (1998) or Malkan \& Stecker (1998), but we 
 specifically address redshift evolution of the MRF and the effects of dust-reemission in 
 the infrared, the initial mass function (IMF) and metallicity. 
 In $\S$3 we discuss the use of population synthesis models to relate the SFR to the
 observed emissivities, and describe models of the dusty ISM in star forming regions
 that allow us to reproduce the far-infrared bump in the present-day MRF spectrum .
 Despite the complexity of the underlying physics involved in the production of the MRF, 
 one can successfully model the MRF with simple modules. This approach allows us to  
 investigate with clarity the various factors contributing to the MRF. In 
 $\S$4 we present the MRF spectrum as a function of redshift and discuss in detail
 the dependencies on cosmological models. We discuss the effect on the IR peak induced
 by varying assumptions about the IMF, the mean metallicity of the emitting stars and the
 effect by adding a new dusty population of galaxies, ULIGs/LIGs (ultraluminous/luminous infrared
galaxies).
 Note that the MRF determined in this way does not depend on the parameters of the 
 assumed cosmological model. However, when we refer to comoving emissivities or
 the cosmic star formation rate SFR(z), we adopt the flat Friedmann model with $\Omega_0=1$,
 $\Omega_\Lambda=0$, and $h=0.5$ where $h=H_0/(100~\rm km~s^{-1}~Mpc^{-1})$. This choice
 of parameters is most commonly made in the observational literature, so we employ it 
 here to allow direct comparisons.

 \section{Method}
 \label{sec:method}
 The method for calculating the MRF from a given SFR relies on an accurate knowledge of
 evolving stellar spectra and the reprocessing of star light in various dusty environments.
 Luminosity evolution of stellar populations is sensitive to the initial mass function (IMF),
 evolution of the mean cosmic metallicity, and the amount of interstellar extinction.
 Starting point of any model is the spectral energy density (SED) produced by a population 
 of stars resulting from an instantaneous burst of star formation (commonly
 normalized to the mass of stars formed). Because star formation is an ongoing process
 with relatively short time scales of 10$^{5-7}$ yrs, the starburst spectra can be directly 
 convolved with the global SFR, $\dot{\rho_\ast}(z)$, to derive the evolution of the global 
 luminosity density due to cosmic star formation. The SEDs are 
 constructed from realistic stellar evolution tracks combined with detailed atmospheric
 models (e.g., Bruzual $\&$ Charlot 1993). The temporal evolution of the specific 
 luminosity, $L_\nu(t)$ (in units of erg~s$^{-1}$Hz$^{-1}$ per unit mass of stars formed)
 is then determined by the choices of IMF and the initial stellar metallicity. 
 Fig.~1 shows the results for a Salpeter IMF between 0.1 and 100 M$_\odot$ 
 and solar metallicity. Note that the figure shows $L_\lambda$ as a function of 
 wavelength. The luminosity drops rapidly as massive stars become supernovae
 (whose light is not included in these SEDs), and the wavelength of the bulk of the 
 emission shifts to the red as the population ages. SEDs shown in Fig.~1 are unobscured 
 by circumstellar gas and dust. The effects of absorption are discussed in $\S$ 3.2.\\ 
 
 From the population synthesis starburst models we obtain the comoving emissivity (or luminosity 
 density) at cosmic epoch $t$ from the convolution
 \begin{equation}
 \mathcal{E}_{\nu}(t) = \int_{t_m}^{t}
 L_{\nu}(t-t')\dot{\tilde\rho}_{\ast}(t') dt' ~ (\mathrm{erg s^{-1} Hz^{-1} Mpc^{-1}}) ~
 \label{eq:emist} 
 \end{equation}
 where $\dot{\tilde\rho}_\ast(t)=\dot{\rho}_\ast(z)$ is the star formation rate per comoving 
 unit volume. Rewriting Eq.~(\ref{eq:emist}) in terms of redshift, $z=z(t)$, yields
 \begin{equation}
 \mathcal{E}_{\nu}(z) = \int_z^{z_{m}}
 L_{\nu}(t(z)-t(z'))\dot{\rho}_{\ast}(z') \left | \frac{dt'}{dz'} \right |
 dz' ~,
 \label{eq:emislambda} 
 \end{equation} 
 where we assumed that star formation began at some finite epoch
 $z_m=z(t_m)$. For given evolution of the emissivity a second integration over
 redshift yields the energy density, or, after multiplication with $c/4\pi$, the
 comoving power spectrum of the MRF
 \begin{equation} P_\nu(z) = \nu
 I_{\nu}(z) = \nu \frac{c}{4\pi} \int_z^{z_m}  \mathcal{E}_{\nu'}(z')
\left | \frac{dt'}{dz'} \right |  dz' ~ , \label{eq:hinter} \end{equation} 
 with $\nu'=\nu(1+z')/(1+z)$.  
 Cosmological parameters enter through $dt/dz$,
 which is given by (e.g., Peebles 1993 \emph{Principles of Physical Cosmology}) 
 \begin{equation}
\left | \frac{dt}{dz} \right | =\frac{1}{H_0(1+z)E(z)}
 \end{equation}
 with an ``equation of state'' 
 \begin{equation}
 E(z)^2= 
 \Omega_{\rm r}
 (1+z)^4+\Omega(1+z)^3+\Omega_R(1+z)^2+\Omega_\Lambda ~.
 \end{equation}
 The term proportional to $\Omega_{\rm r}$ takes into account the contribution from relativistic
 components (such as the CMB and the star light). The density parameter of this component is
 defined as $\Omega_{\rm r} =
 u_{\rm r}/\rho_{\rm crit}c^2$, where
 $u_{\rm r}$ refers to the relativistic energy density and $\rho_{\rm crit}$ is the 
 critical density of the universe; $\rho_{\rm crit}$ = 3H$_0^2$/8$\pi$G = 10.54 h$^2$ keV/cm$^3$.
 The resulting dependencies on the Hubble constant are;
 emissivity $\propto H_0^{-1}$, and MRF power spectrum $\propto H_0^{-2}$. However,
 this scaling is only correct if we formally require that the star formation rate is a
 given function of redshift. Lacking a reliable derivation from
 first principles, this function is derived from observations that
 involve distance and luminosity estimates, which introduces additional powers of H$_0$.  
 
 \section{Emissivity}

 \begin{figure*}
\centering
  \includegraphics[width=12cm]{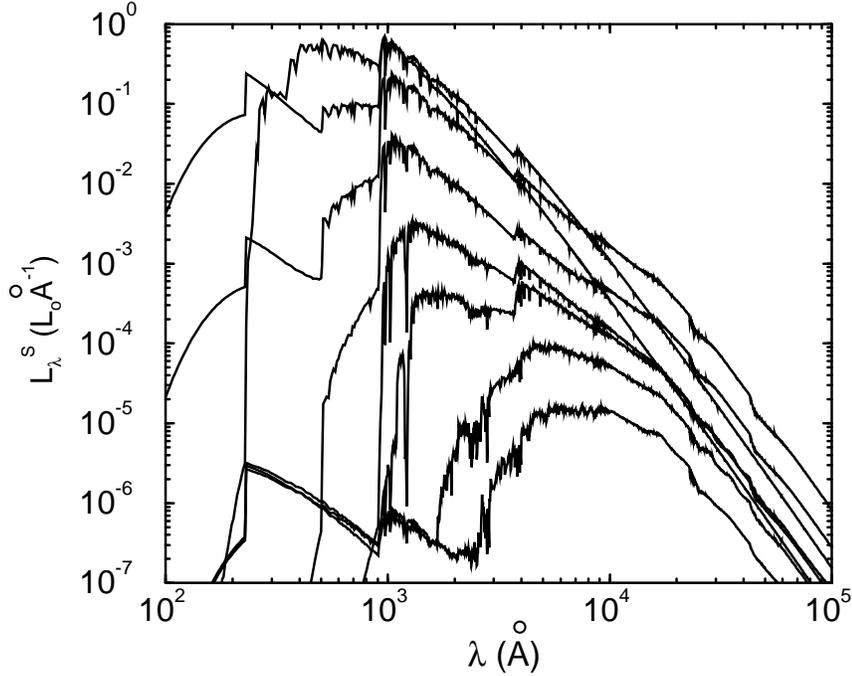}
  \caption{Spectral energy distributions (SEDs) of a coeval stellar population as a function
 of age $\tau$. The specific luminosity of the evolving stellar population is normalized to one
 solar mass. The stellar models assume standard solar composition. Shown are the
 SEDs for ages $\tau$ 0, 0.0038, 0.00724, 0.0138, 0.07187, 0.28612, 1.434, and 11 Gyrs (from top  
 to bottom, based on calculations by Bruzual \& Charlot 1999).}
    \label{fig1}
 \end{figure*}

 \subsection{Spectral synthesis model}
 Conversion of gas to stars produces a stellar mass distribution that is commonly described 
 by a ``universal'' initial mass function. A batch of stars produced in an instantaneous
 ``burst'' of star formation is often referred to as a Simple Stellar Population (SSP). Massive 
 stars in the SSP have short lives ($\sim$ 10$^7$ yrs) and predominantly produce UV radiation,
 while long-lived, low-mass stars remain close to the main sequence even over cosmological 
 times and produce the bulk of the ``red light''. Depending on their mass, stars
 follow different evolutionary paths and evolve on different time scales, which causes the SED
 of a SSP to be a sensitive function of time. To follow the changing energy output in time we  
 use the population synthesis code of
 Bruzual \& Charlot 1999 (BC-Model), which is an updated version of the code
 documented in GISSEL96 (Leitherer et al. 1996). Figure 1 shows the resulting SEDs emitted
 at several distinct times after the burst. The stellar spectra used to construct these SEDs 
 are based on Padova tracks (e.g. Girardi et al. 2000) and Lejeune stellar atmosphere models 
 (Lejeune et al. 1997, 1998), and include a post-AGB evolutionary phase. For demonstration purposes 
 we adopted a population with solar metallicity and Salpeter single power law IMF in the range
 0.1$M_\odot < M < 100 M_\odot$,
 although different choices can readily be made with this code. The SEDs shown
 in Fig.~(\ref{fig1}) are of course not representative of any particular type of
 galaxy, because those involve star formation histories that are usually different from
 a single burst. However, the SEDs do resemble more closely the spectra of elliptical 
 galaxies, for which the single star burst might be a reasonable approximation.\\

 \subsection{Interstellar Medium}

 \begin{figure*}
\centering
\includegraphics[width=12cm]{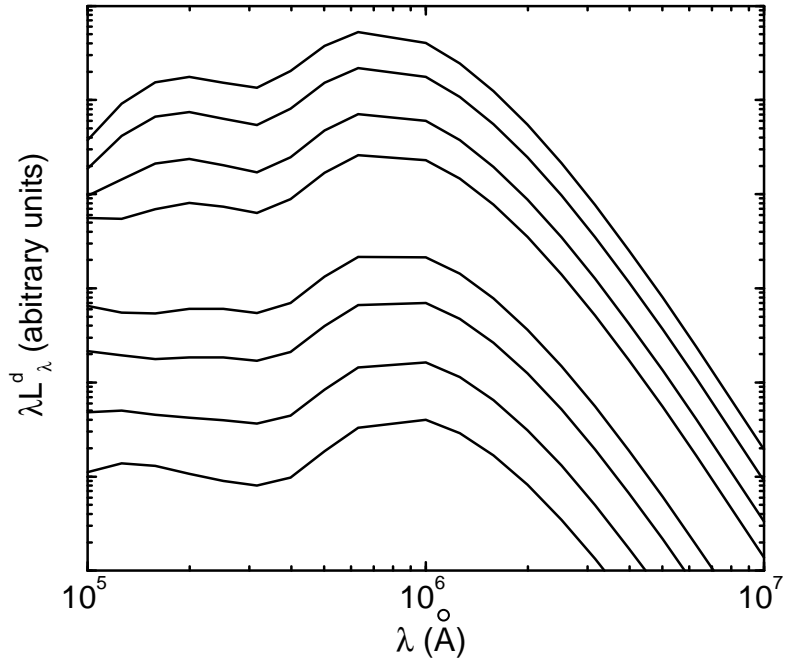}
 \caption{IR-spectra associated with each of the SSP-spectra shown in Fig~\ref{fig1}.
 The IR emission is modeled as the sum of three modified blackbody spectra. The  determination of the temperatures of these components is discussed in the text.}
 \label{fig2} 
 \end{figure*}
 
 In contrast to the intergalactic medium (IGM), which is extremely tenuous and has only 
 a small effect on the transmitted SEDs from redshifts less than unity (e.g., Madau
 2000), the intrinsic absorption by the galactic interstellar
 medium (ISM) is significant, and must be included to obtain correct SEDs. For
 simplicity, we assume a uniform distribution of gas and dust surrounding the stars
 of the SSP, and apply Osterbrook's Case B recombination for optically thick clouds 
 at an equilibrium temperature of $10^4$K, i.e. total absorption of all ionizing photons
 and reemission of 68\% of the absorbed power in $L_\alpha$ line emission (which
 is subsequently absorbed by dust). The remaining energy is assumed to be reemitted in 
 the optical regime via bremsstrahlung and recombination line/continuum
 emission. The assumption of homogeneity of the absorbing gas and dust layers is a  
 simplification that could have a noticeable effect on the estimated transmission of 
 UV radiation shortward of the Lyman edge. A more realistic approach should also take
 into account ionized superbubbles that are driven into the ISM by multiple supernovae 
 (e.g., Dove, Shull, $\&$ Ferrara 2000).
 
 The average metallicty of gas in galaxies slowly increases with cosmic time, but
 the present-day value is not known precisely (e.g., Pei, Fall, $\&$ Hauser 1999).
 We thus adopt an average extinction curve 
 \begin{equation} 
 A_\lambda=0.68\cdot E(B-V)\cdot R\cdot (\lambda^{-1}-0.35)
 \end{equation}
 with $R=3.2$ and
 where $A_\lambda$ with $\lambda$ [$\mu$m] determines the absorption coefficient according to
 $g(\lambda)=10^{-0.4\cdot A_{\lambda}}$.
 Reemission is calculated as the sum of three modified Planck spectra

 \begin{equation}
 L_{\lambda}^d(L_{bol}) = \sum_{i=1}^{3} c_i(L_{bol})\cdot Q_\lambda \cdot B_\lambda(T_i)
 \label{eq:planck} \end{equation}
where

$Q_\lambda \propto \lambda^{-1}$ and $L_{bol} = L_{bol}(\tau)$. 

 Two temperatures characterize warm and cold dust in galaxies. The third temperature is 
 included to model a PAH component, which is assumed to emit like a Blackbody. 
 Emission lines are not treated separately, because of the smoothing effects from 
 integration over redshift. We normalize these three components realtive to each other
 (with the constants $c_i, i=1\ldots3$) by using non-Seyfert galaxy observations by Spinoglio et al. 
 (1995). We fitted the relation for all 4 bands.

 Dust in the ISM of the Milky Way is known to coexist at several different
 temperatures, determined by the distances from various heat sources. Hot dust in spiral
 galaxies has temperatures ranging 
 from 50~K to 150~K-200~K (Sauvage {\em  et al.} 1997 and references therein) 
 when in thermal equilibrium
 with HII regions, young massive stars, or compact accreting sources. Radiation from
 this dust component 
 emerges in the mid-infrared and reprocesses only a small fraction of the
 emitted luminosity.  Warm dust with temperatures between 25~K and 50~K corresponds to
 regions heated by the mean interstellar radiation field. Dust inside molecular clouds
 is somewhat shielded against high-energy radiation, and thus appears at low
 temperatures between 10~K and 25~K. Very cold dust at temperatures of 10~K, or even 
 less, can be present in the densest parts of molecular clouds or in outer regions of
 the galaxy where the flux of the interstellar radiation field has dropped to
 the value of the intergalactic radiation field. Such very cold dust is
 difficult to detect, and requires sub-mm observations which so far have failed
 to provide unambiguous results. Therefore, we do not include very cold 
 dust in our model. To keep the model simple, we consider variations in dust
 composition only because the shape of the spectrum is dependent of the total flux. 
 It is noteworthy, however, that emission features around 10~$\mu$m 
 due to Polycyclic Aromatic Hydrogen (PAH) molecules seem to be ubiquitous
 in galaxies (Desert el al. 1990). These PAH's are undergoing
 temperature fluctuations and are generally not in thermal equilibrium. 
 The broad emission lines of the PAHs are modeled with an additional 
 (low flux) blackbody component, characterized by
 T $\sim 425$~K (Dwek et al. 1997). \\

 As mentioned above, we use the non-Seyfert galaxy relations for all 4 bands by Spinoglio et al. 
 (1995) to fix the 6 model parameters in Eq.~\ref{eq:planck}
 (three temperatures $T_i$, i=1...3 and three constants $c_i$, i=1...3). 
 The relations depend on the 
 total luminosity which is radiated by a galaxy. 
 Although we are using a SSP and not a ''real'' galaxy we can use the IR relations of galaxies, 
 because these relations 
 only depend on the total luminosity of a galaxy.   We use $10^{11}$M$_\odot$ for each SSP 
 spectrum to get a galaxy-like total luminosity. Note that the dust-spectra $L_\lambda^d$ 
 depend on the age $\tau$ of the SSP, because they change with total luminosity $L_{bol}$.

 As starting values for the temperature we use
 $T_1$=240K (consistent with a PAH blackbody of 425K), $T_2$=80K (warm dust) and $T_3$=30K
 (cold dust). After the fitting procedure we obtain for each SSP spectrum the associated
 IR-spectrum (see Fig.~2).
 The resulting temperatures are higher for higher total luminosities (for younger SSPs) 
 and they are generally in the range 27K$<T_1<$33K, 125K$<T_2<$70K, and 180K$<T_3<$400K.

 The resulting total spectra, including absorption
 and reemission, can be written for each SSP-spectrum as
 \begin{equation}
 L_{\lambda}(\tau)=[(L_{\lambda,{\rm
 BC}}(\tau)+c_g(\tau)\cdot \epsilon_\lambda)]\cdot g_{\lambda}+c_d\cdot L_{\lambda}^d(\tau)
 \label{eq:spekgal} \end{equation}
 where $L_{\lambda,{\rm BC}}(\tau)$ are the SEDs from the BC-Model with an age $\tau$. 
 The optical emission of gas
 heated by absorbed photons is included through the quantity 
 $\epsilon_{\lambda}\ \propto {\rm exp}(-h\nu/kT)$
 multiplied by the fractional energy $c_g$ available for this channel (32\%). Both
 spectral components are modified by the absorption coefficient $g_\lambda$. The
 dust reemission spectra $L_{\lambda}^d(\tau)$ are added according to
 Eq.~(\ref{eq:planck}). The IR-spectra shown in Fig.~\ref{fig2} are normalized with 
 $c_d$ using energy conservation of the absorbed and re-emitted photons.\\
 
 Published values for the color index, $E(B-V)$, cover a wide range, which 
 reflects large uncertainties associated with dust properties. Steidel et al.
 (1999) adopt $E(B-V)=0.15$ at redshifts $z=3-4$. Madau et al. (1998) use a
 universal value of $E(B-V)=0.1$, Guiderdoni (1999) prefers $E(B-V)=0.09$ at
 $z>2$. Generally, dust extinction seems to play a more important role at high
 redshifts (e.g., Pettini et al. 1998). According to Madau et al (1998) the index
 varies as $E(B-V)=0.011(1+z)^{2.2}$.

 Using the relation $\lambda L_\lambda(\tau)=\nu L_\nu(\tau)$ we calculate $L_\nu$ 
 to obtain the emissivity in Eq.~\ref{eq:emislambda}

 \subsection{Star formation history}
 \begin{figure*}
\includegraphics[width=9cm]{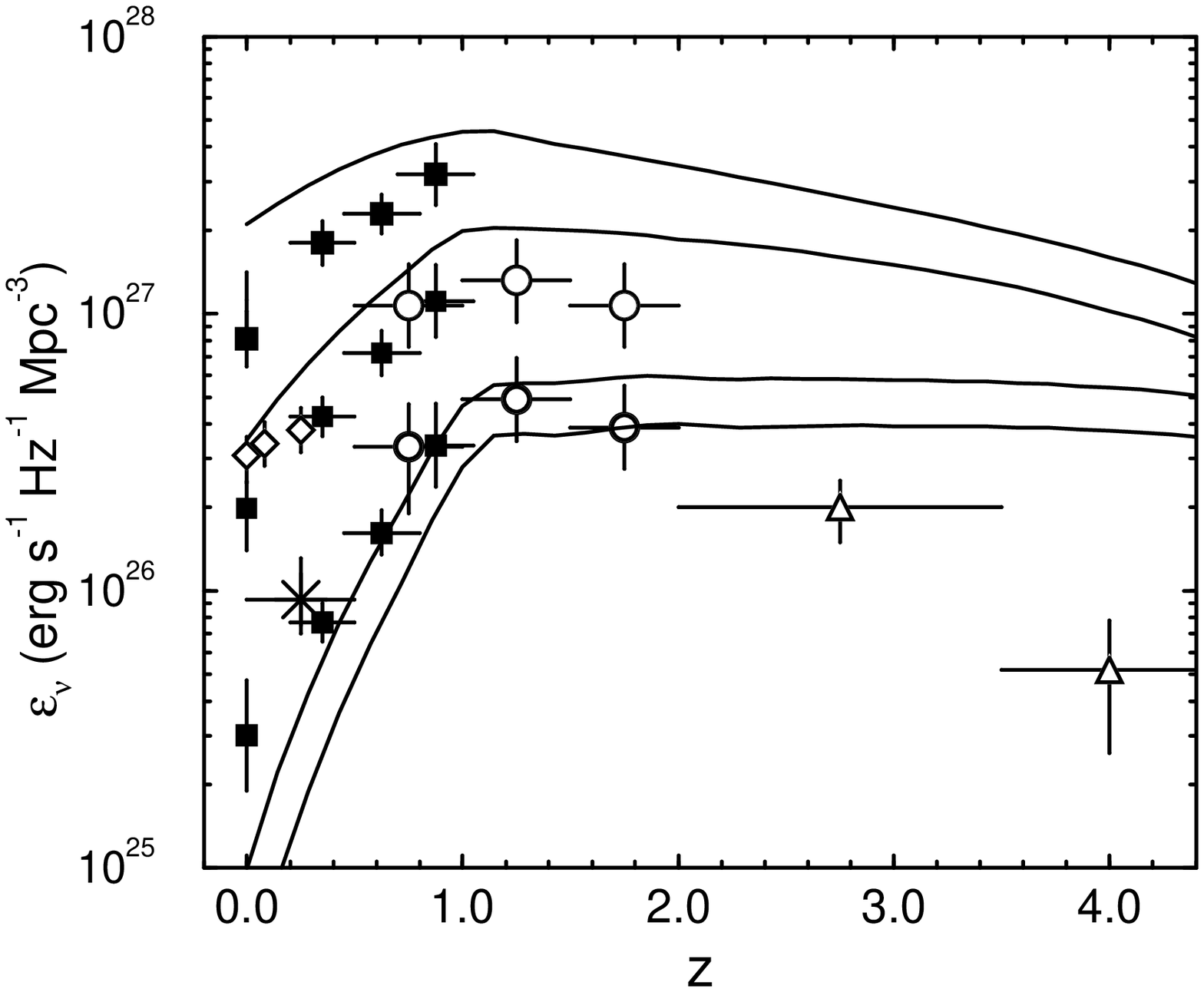}
\includegraphics[width=9cm]{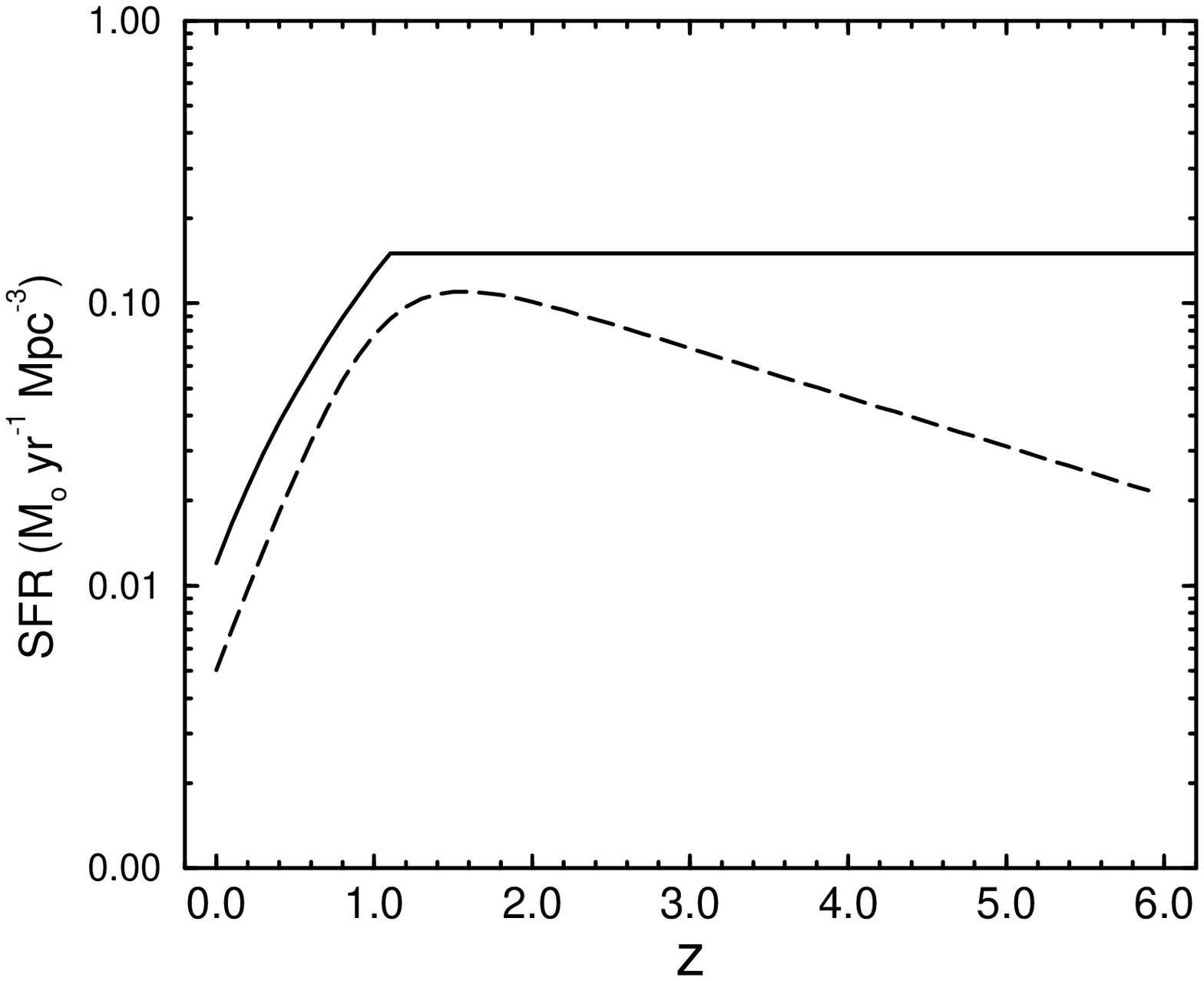}
 \caption[fig3a.eps]{a)
 Redshift dependence of the emissivity corresponding to the star formation history 
 shown in the right panel. {\it Solid lines} show model emissivities at 1.0~$\mu$m,
 0.44~$\mu$m, 0.28~$\mu$m  and 0.16~$\mu$m from top to bottom. Data plotted with
 {\it solid squares} are taken from Lilly et al. (1996), {\it open circles:} 
 Conolly et al. (1997), {\it solid diamonds:} Ellis et
 al. (1996) and {\it open triangles:} Pozzetti et al. (1998; 
 lower limits at high redshift and  0.16$\mu$m).
 b) Comoving star formation rate density as a function
 of redshift. The solid line is the rate used for computing the emissivity
 (with $\alpha=3.4$, $\beta=0$, $z_{peak}=1.1$,
 $\dot{\rho_\ast}(z_{peak})=0.15~\rm M_\odot~yr^{-1}~Mpc^{-3}$. 
 The dashed line is a fit function provided by
 Madau (1999). \label{fig3} }
 \end{figure*}

 Recent deep galaxy surveys (e.g., Kennicutt 1983, 1999) or Lyman $\alpha$
 absorber studies (Pei \& Fall 1995; Pei, Fall, $\&$ Hauser 1999) suggest a 
 functional shape of the
 SFR ($\dot{\rho_\ast}(z)$) that can be approximated with a simple
 broken power law 
 \begin{equation} \dot{\rho_\ast}(z) \propto (1+z)^\alpha \end{equation} 
 with $\alpha=\alpha_m>0$ for $z \le z_{peak}$ and $\alpha=\beta_m<0$ for $z >z_{peak}$.  
 Plotted as a function of redshift the right hand panel of Fig.~\ref{fig3} 
 shows the fit function in comparison to the more complex fit function
 given in Madau (1999).
 The cosmic star formation rate density SFR(z) has been determined with
 different methods and for large set of input data, as recently summarized
 by Ruiz-Lapuente et al. (2000). These studies suggest that the
 original Madau curve, Madau (1997(II)), should be considered a lower limit, 
 and that realistic rates could be
 larger by a factor 2$-$3 at all redshifts. The compilation of
 Ruiz-Lapuente et al. (2000) (see their Fig.~1) clearly shows that we 
 do not yet understand systematic effects well enough to obtain a reliable
 estimate for SFR(z). This is especially true at redshifts much beyond unity.
 
 In our approach to modeling the MRF, the SFR function is 
 considered to be a free fit function aimed at reproducing the emissivities
 derived from deep surveys. "Measurements" of the SFR are generally based on 
 luminosity densities, and thus model dependent. It is thus preferable to use
 the emissivities directly to obtain a self-consistent star formation history. 
 We note, that the SFR parameterization used here does not contain any
 cosmological parameters. However, it is clear that choosing a different
 cosmology does change the observationally determined emissivities (Lilly et al.
 1996, Ellis et al. 1996), hence requiring a different SFR.\\
 
 For a given SFR, the emissivity is readily obtained from the convolution
 given by Eq.~\ref{eq:emislambda}.
 The resulting $\mathcal{E}_{\nu}(z)$ is plotted in Fig.~\ref{fig3} for
 four different wavelengths in the optical band, and compared to the 
 observations. Note that the steep increase at 0.28~$\mu$m and
 the shallower increase at 1.0~$\mu$m are reproduced by the model.
 The model slightly overproduces the present-day emissivity at 1.0~$\mu$m. 
 However, the data point
 of Lilly et al. (1996) at 0.44~$\mu$m and $z=0$ falls much below the
 corresponding value obtained by Ellis et al. (1996), indicating the conservative
 nature of the Lilly et al. measurements. Generally, there is good agreement
 with the data.

 \subsection{Contribution from luminous infrared galaxies }
 Luminous Infrared Galaxies (LIGs) were first discovered with IRAS (Soifer et al. 1987).
They represent a population of galaxies with IR luminosites above $10^{11}$L$_\odot$ 
 (L$>10^{12}$L$_\odot$ are named ultraluminous infrared galaxies ULIGs) and high 
 star formation rates. Most of these galaxies are dust enshrouded starburst galaxies or
 mergers, some of the ULIGs have also been identified as AGN
 (e.g. Kim et al. 1998).
 Although the LIGs are not so numerous today, a significant fraction of the infrared light
 could originate from them.  Moreover, their star formation history could well be different from
the SFR of optical galaxies (Franceschini et al. 2001). 
As these galaxies do hardly show up in optical surveys (E(B-V)$_\mathrm{LIG}>$ E(B-V)$_\mathrm{OPT}$), there
existence is not reflected in the SFR discussed in the previous section. Also, the dust temperatures
in LIGs are higher,
Hence we emphasize that adding another SFR$_\mathrm{LIG}$ is a straightforward ramification of the model
which would affect the infrared part of the MRF.
 In the following we will start to model the MRF without this contribution,
 but we will return to it later.

 \section{Results}

 \subsection{Metagalactic radiation field}
\begin{figure*}
\centering
\includegraphics[width=13cm]{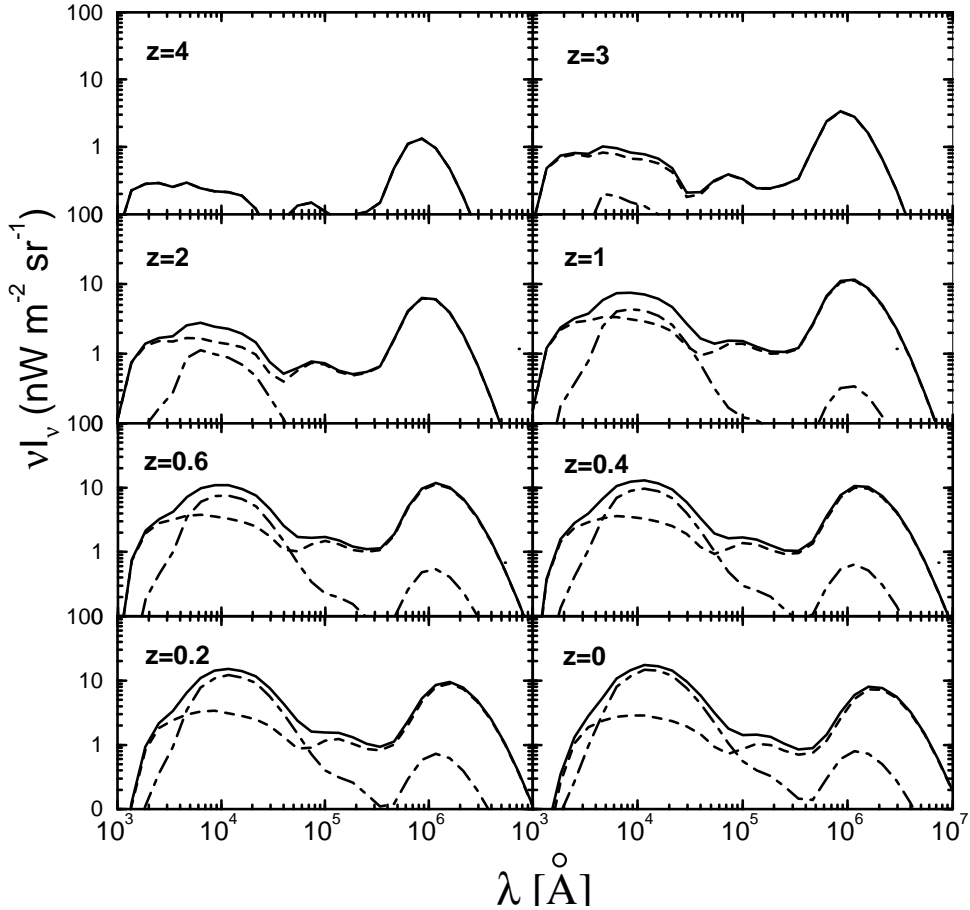}
 \caption{The evolving spectrum of the extragalactic background light
 (with model parameters as
 in Fig.~\ref{fig3}) as a function of wavelength (in Angstroms). 
 The {\em dashed lines} show the contribution of massive stars (MS life $< 0.3Gyr$) and the 
 {\em dot-dashed line} the contribution of low mass stars (MS life $> 0.3Gyr$).\label{fig4}}
 \end{figure*}

 The final step in computing the MRF involves an integration of the
 emissivity over cosmic time using Eq.~(\ref{eq:hinter}),
 where we neglected spectral modifications due to the intergalactic medium
 (IGM). The IGM consists predominantly of Lyman-$\alpha$ clouds
 with HI, HII and HeI gas (e.g., Madau 1997(I)) and mainly affects photons 
 with wavelengths shortward of 911\,{\AA} which we assumed to be completely
 absorbed inside galaxies. Recent work by Pei, Fall \& Hauser (1999) also
 shows that absorption in the IGM, if present, is minimal. The neglect of IGM
 effects can cause a slight underestimate of the star formation rate, but 
 will not affect the estimate of the MRF spectrum, because our fit procedure 
 uses the observed emissivities, which guarantees that all photons contributing
 to the MRF are accounted for. The evolution of the comoving MRF spectrum is shown 
 in Fig.~\ref{fig4} for several redshifts, where the wavelength scale corresponds 
 to a comoving reference frame. 
 
 \subsection{Dependence on cosmology}
 
 Formally, the MRF intensity computed according to the procedures outlined in
 Sec.~\ref{sec:method} exhibits an explicit cosmology dependence. This is the 
 consequence of choosing an apparently cosmology-independent SFR. Such a choice 
 is unrealistic, because if one were to change the cosmological
 model, while keeping the same SFR,
 one would immediately fail to reproduce the observed emissivities. However, the 
 emissivities themselves are computed directly from observables, viz. the fluxes
 $\mathcal{F}$ and redshifts of galaxies. We therefore have to take a look
 at the cosmology dependence of the emissivities implied by the relation
 \begin{equation} \mathcal{E}(z) =  \frac{dL}{dV_c}=\frac{4\pi
 d^2_L(z)d\mathcal{F}}{dV_c} \label{eq:emidat} \end{equation} where $d_L(z)$ is
 the luminosity distance and  \begin{equation} dV_c=\frac{d_L^2\,
 d\Omega\, dz}{H_0(1+z)\,E(z)} \end{equation} the comoving volume element,
 and hence $\mathcal{E}(z)\propto H_0 (1+z) E(z)$. Inserting this in
 Eq.~(\ref{eq:hinter}) strictly cancels the cosmology dependence. This argument
 means that by requiring the model to fit measured emissivities (evaluated for a
 given cosmology) the resulting MRF no longer explicitely depends on the choice
 of cosmological parameters. The background radiation field becomes, in a
 sense, a measured quantity itself, since both luminosity and volume scale
 $\propto d_l^2$.  \\
 
 \subsection{Bolometric flux}
 We complete this section with a comparison of the ``bolometric'' (IR - opt) flux obtained
 from our model with results available in the literature (Table~\ref{tab2}). A strict
 lower limit on the present-day MRF flux of 28~nW m$^{-2}$ sr$^{-1}$ was derived from COBE 
 and HST data (Dwek et al. 1998). 
 The integrated flux from our model, 46~nW m$^{-2}$ sr$^{-1}$, is in agreement with all 
 models using a similar SFR.
 
 \begin{table}
\centering
\caption{Integrated diffuse background for different models.}
 
 \begin{tabular}{ll}
 \hline \hline\
 & $I(0)$ [nW m$^{-2}$ sr$^{-1}$] \\
 \hline
 This Model &  46\footnotemark\\ 
Range from Data & 55 $\pm$ 20\footnotemark\\
Dwek et al.(1998) (UVO)\footnotemark & 30   \\
 ..............................(PFI)\footnotemark & 91 \\
 ..............................(PFC)\footnotemark  & 41    \\
 Pei, Fall \& Hauser (1998) & 51-55  \\ \hline                  
 \end{tabular}
 \label{tab2}
 \end{table}
 \footnotetext[1]{I(0.2)=45; I(0.4)=44; I(0.6)=42; I(1.0)=35; I(2.0)=15; I(3.0)=7; I(4.0)=2}
 \footnotetext[2]{Pozzetti\&Madau (2000)}
 \footnotetext[3]{using SFR-Madau et al.(1998)}
 \footnotetext[4]{using SFR-infall model (Pei\&Fall (1995))}
 \footnotetext[5]{using SFR-closed box model(Pei\&Fall (1995))}

  \section{Discussion}

 We have developed a simple model and its evolution with time. This model is
 based on direct measurements of the global emissivities due to galaxies.
 There are no measurements of the MRF at high redshifts that could be
 used to directly verify our model, with the possible exception of the UV
 background based on the proximity effect (e.g., Giallongo et al. 1996). 

\begin{figure*}
\centering
\includegraphics[width=12cm]{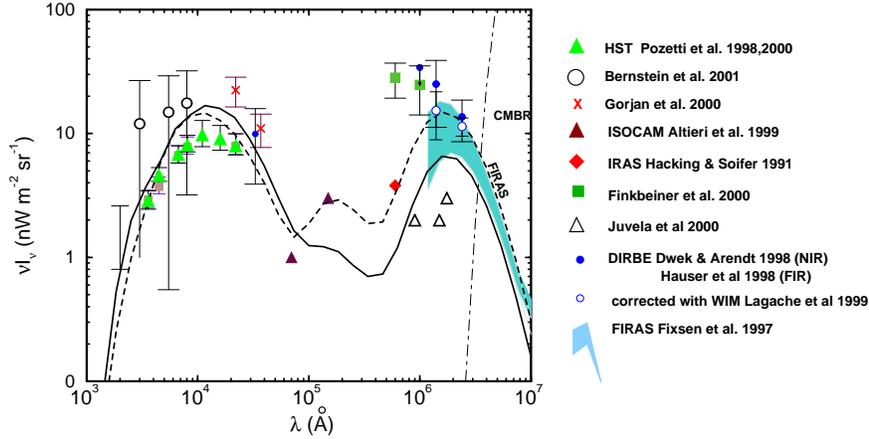} 
\caption{The EBL computed with the model (solid line).  Parameters are: Z=0.02, 
E(B-V)=0.14 (young population), E(B-V)=0.03 (old population), SFR-parameters see Fig3.
The deficit in the far-infrared can be avoided using a (perhaps unrealistically)
low value of the metallicity (Z=0.0001).  The effect of low-Z stellar atmospheres is to produce relatively
more UV radiation which is subsequently redistributed to the FIR by interstellar dust grains.
\label{fig6}}                  
\end{figure*}

 In the second paper of this series we will introduce a method to constrain the MRF
 at high redshift using this model and high-energy observations of blazars.
 For now this leaves us with various measurements of the present-day
 MRF as the most relevant set of constraints. While the match is by no means
 perfect, the comparison shown in Fig.~\ref{fig6} suggests that our simple
 model is capable, without fine-tuning of parameters, of explaining the global 
 level and the general spectral shape of the MRF. But the predicted flux in the
 IR-band falls short by roughly a factor two. As discussed above, the magnitude
 as well as the shape of the MRF are the result of a convolution of SEDs from
 an aging stellar population with continuous star formation at a given rate
 SFR(z) and the cosmological effects of redshift and time dilation. The fact that
 our model roughly matches both magnitude and shape of the MRF indicates that the
 relevant input physics has been properly taken into account. This provides 
 confidence in the predictions of the MRF at higher redshifts, which is needed 
 for a variety of astrophysical studies. However, there are still some deviations that 
 need to be addressed, especially the shortfall in the IR band.\\

\subsection{Population synthesis models}

 There are several model parameters that introduce significant uncertainties in the
 estimated MRF flux. Most important among them are the IMF, the dust extinction model,
 the parameterization of the dust emission by multiple blackbodies with different
 temperatures, the metallicity dependence of stellar evolution tracks, and the 
 amplitude and form of the SFR (especially at redshifts less than unity).
 The choice of cosmology plays only a minor role for the MRF
 spectrum, as we emphasized in the previous section. 
 We also investigated different
 population synthesis templates for the stellar light output (e.g., Leitherer et al. 1996 
 and references therein) and found them to be very similar to each other. The choice 
 of the population synthesis model thus does not significantly affect the estimated MRF.\\ 
 
\subsection{IMF}

 We proceed with a discussion of those parameters that, at least in principle,
 can significantly alter the estimated intensity, flux, or energy density.
 We start by comparing results based on the Salpeter and Scalo IMF (see Fig.~\ref{fig6}. 
 One of the distinguishing features of the Scalo IMF is the fact that it contains 
 relatively few high-mass stars. These stars are 
 responsible for most of the UV photons, which, after thermalization by the dust
 in the ISM, emerge in the FIR. On the other hand, the Scalo IMF contains a relatively
 large fraction of low mass stars which emit most of their light at optical wavelengths.
 Consequently the present-day MRF has a much
 weaker FIR bump when the Scalo IMF is employed. This perhaps provides an argument in 
 favor of the Salpeter form for a global IMF, because the Scalo IMF would somewhat 
 underproduce the FIR background. \\

\subsection{Star formation rate}
 The star formation rate density obtained here (see Fig.~\ref{fig3}) is higher than
 the SFR originally suggested by Madau (1997(II), 1999), and is somewhat
 different from the rate derived from structure-formation theory  
 (Primack et al. 1998). As discussed above, recent determinations of
 the cosmic star formation history based on H$_\alpha$ emission and ISO data
 suggest that the Madau rate has been systematically underestimated by a factor
 2$-$3 (see Flores et al. 1999 for a recent discussion). The measurements
 of the sub-mm SCUBA array (Hughes et al. (1998) support the notion that
 much of the star formation activity at high redshifts is hidden due to dust
 absorption. Ruiz-Lapuente, Casse, $\&$ Vangioni-Flam (2000) summarize many of
 these measurements and compare (their Fig.~1) the various 
 functions with that of Madau et al. (1998). The function used in this
 study falls above the "Madau curve", but below most of the curves compiled
 by Ruiz-Lapuente et al. (2000).

\subsection{Absorption and re-emission of the ISM}

 The dust and gas model we use is not based on first principles, but is
 founded on empirical results. A three temperature model for galactic
 dust spectra  has been proposed by Dwek et al. (1998). A small change
 in the temperatures (say, $\pm 10$K) would only cause a small change in the
 shape of the MRF spectrum. This lack of sensitivity originates from the broadening 
 of the employed modified blackbody spectra due to the integration over redshift. 
 
 Any modification of $E(B-V)$ changes the spectral shape of the MRF from 911\,\AA \,to 
 $10^4$\,\AA, especially the amplitude of the far-infrared bump. This is simply due
 to energy conservation. An increase in the extinction causes a larger fraction of  
 UV absorption, and this energy re-emerges predominantly in the FIR. We selected 
 a value of $E(B-V)$ that provided an acceptable fit to the available data on 
 emissivities as a function of redshift and the present-day background. We found 
 that $E(B-V)=0.14$ is the appropriate value for young SSP's and 0.03 for old SSP. 
 These values are reasonably well determined by the emissivity-fit alone. In any case, 
 changing the $E(B-V)$ value does not provide a solution to the "missing IR flux".

\subsection{Comparison with data and other models at $z=0$}

 The model present-day MRF flux at optical wavelengths is consistent with lower limits from
 HST (Pozzetti et al. 1998, 2000). 
 Observations by Bernstein et al. (2000) suggest 
 the possibility of a somewhat higher MRF flux, but our results still fall within
 their estimated uncertainties. 

 Absolute measurements (albeit
 with large systematic errors due to the need for subtraction of dominant 
 local foreground emission Lagache et al. 1999, Puget \& Lagache 2000) are available from COBE/DIRBE and FIRAS in the IR band.
 A discrepancy between our model and the observations occurs at $\mu$m wavelengths,
 where the measured flux appears to be larger by a factor of two.
 A lower limit based on galaxy counts due to Elbaz et al (1999) at  
 $\sim 10^5$\AA~lies in the range of possible PAH emission.
 
 While the implications for the MFR at high redshifts are rarely stated in the literature, 
 considerable effort has gone into the calculation of the present-day MFR, i.e., the
 extragalactic background light (EBL). In spite of great diversity in the
 computational approaches employed (see the discussion in the introduction), the models generally
 show the same IR deficit noticed in this study (e.g. Dwek et al 1998, Primack 1998).
 Dwek et al.~ suggested a possible solution by adding a new, distinct component of obscured 
 galaxies which emit preferentially in the IR band.

\begin{figure*}
\centering
\includegraphics[width=12cm]{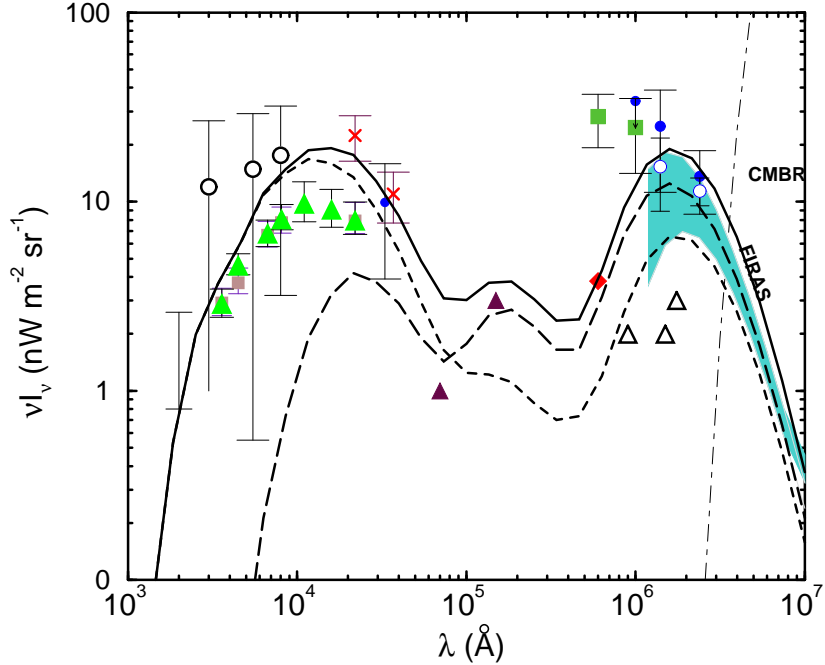} 
\caption{The EBL computed with the model including a separate LIG population (solid line), for the same
parameters that were used to produce the solid line in Fig.~5. 
The long dashed
line shows the contribution from LIGs (parameters see text 5.7), the short dashed line the contribution from
optical galaxies. 
\label{fig7}} 
\end{figure*}

 \subsection{Metallicity}

 The calculations for our "standard model" employ stellar models with solar metallicity. 
 Salamon \& Stecker (1998) 
 and Pei, Fall \& Hauser (1999) discuss some of the possible effects caused by changes 
 in the mean cosmic metallicity. The approximation of a fixed, high (solar) metallicity 
 is motivated in part by structure formation simulations (e.g., Valageas $\&$ Silk 1999)
 which suggest that the mean metallicity in star forming regions is a slowly rising
 function for redshifts less than 2. The calculations of Valageas $\&$ Silk
 also indicate that the metallicity has been larger than 1/2 Z$_\odot$ since z $\sim$ 2 
 (note that their estimate only includes enrichment due to SNII). The significantly lower
 metallicity values found in damped Lyman Alpha systems (DLAs; e.g., Pettini et al.
 1997) probably correspond to the enrichment history of galactic halos. The trend of [Zn/H]
 with redshift suggests a present-day metallicity of 1/3 solar (Vladilo et al. 2000),
 which also indicates that DLAs do not trace the chemical evolution of proto-disks, but
 instead star-forming fragments which build up galaxies through mergers. Although the 
 Milky Way does not represent a good template for cosmic chemical evolution (e.g., 
 Prantzos $\&$ Silk 1989; Fields 1999) its age-metallicity relation and well known
 radial metallicity gradients suggest that much of its current star forming activity
 takes place in environments with Z $\sim$ Z$_\odot$, or even higher (see Boissier $\&$
 Prantzos 1999 for a recent model of galactic chemical evolution). We thus expect the
 Z = Z$_\odot$ = const assumption to provide a reasonable approximation. However, to test
 the dependence of our results on metallicity, we performed the MRF simulation for two 
 cases: i) solar metallicity Z$_\odot$ = 0.02, and ii) Z = 0.0001. The dashed lines in 
 Fig.~\ref{fig7} show the ``low-Z MRF'' in comparison to the standard model with solar
 metallicity (solid line). Both simulations used the Salpeter IMF. The comparison 
 indicates that metallicity effects could be important.\textrm{ Lower metallicity in the stellar 
 atmospheres leads to a higher fraction of light primarily emitted in the UV, and subsequently
 redistributed towards the IR by interstellar dust grains, with extinction parameters determined
 newly from the fit to the emissivities (i.e. independent of the assumed low metallicity
in the stellar atmospheres producing the bulk of the light).  A rigorous treamtent of
metallicity effects, in order to obtain self-consistent interstellar extinction curves
and chemical evolution (Pei, Fall, \& Hauser 1999), is beyond the scope of the paper.}

\subsection{ULIGs/LIGs}
The approach in our model up to this point was to consider the emission from galaxies
found in deep optical surveys, and to compute a mean galaxy spectrum
for them.  This is certainly a valid scheme to obtain a lower limit to
the MRF, and, in fact, the model does not overproduce observational upper limits
at any frequency.  However, the infrared deficit revealed by the
preceding analysis, seems to justify the
inclusion of a population of luminous infrared galaxies (see Sect. 3.4).
This would
increase the strength of the infrared bump in the MRF, but leave
the optical bump unchanged. Hence the overall performance of the
model to reproduce the present-day extragalactic background spectrum can be improved.
The star formation history used for this population
SFR$_\mathrm{LIG}$ (Dwek et al. 1998) has been infered from number counts 
in ISO, IRAS and SCUBA data, and the
luminosity functions from Chary \& Elbaz (2001) 
(and references therein).   
The extinction parameter is set to be E(B-V)=1.0, a value which is typical for 
LIGs. The resulting SFR$_\mathrm{LIG}$ ($\alpha$ = 5.5, $\beta$ = 0, z$_{peak}$=1.0, 
$\dot{\rho_\ast}(z) =0.10~\rm M_\odot~yr^{-1}~Mpc^{-3}$)  is in agreement with the one calculated by Chary \& Elbaz.
At low redshifts the contribution of stars in LIGs is very small, but at high 
redshifts the number of stars formed in LIGs is comparable to the one in optically 
selected galaxies. This is in line with the interpretation of the IRAS (Kim \& Sanders 1998) 
observations, which indicate that LIGs were more numerous 
in the past than they are today.
Fig.~6 shows the EBL and the respective contributions from optical and infrared galaxies.

\section{Conclusions}
We have developed a model for the evolving MRF based on optical galaxy surveys as its
main observational input, and found that this model shows a deficit at infrared
wavelengths in the spectrum of the EBL (the MRF at $z=0$).
Inclusion of obscured, infrared-emitting galaxies provides a viable solution of the problem,
and we have determined their SFR from fitting the model to EBL data.
The model in this form can serve as a reliable basis for obtaining predictions of the MRF at high
redshifts.  Observations of high-redshift gamma ray sources with next-generation gamma ray telescopes
(GLAST, HESS, MAGIC, VERITAS) are expected to soon provide evidence for gamma ray attenuation
due to collisions of gamma rays with low-energy photons from the MRF, thus allowing to
test the model predictions in an independent way (paper II in this series).
\footnote{
At http://wwww.astro.uni-wuerzburg.de/theorie/ we provide a web tool for online 
calculations of the MRF, allowing users to select values of key parameters.}.
 
 \begin{acknowledgement}
 We thank Stephane Charlot for providing the latest release of the population 
 synthesis templates (GISSEL99), and 
 Joel Primack for communicating the SFR function from structure formation
 theory prior to publication.
 We also thank Floyd Stecker for fruitful discussions.
 This work is part of a research project
 supported by the BMBF under code DESY-HS/AM09731M45.
 DH acknowledges support and 
 hospitality during visits to the Universit\"ats-Sternwarte G\"ottingen. 
\end{acknowledgement}

 \end{document}